\documentclass[final,1p,times]{elsarticle} 
\usepackage{graphicx} 
\usepackage{amssymb} 
\usepackage{amsthm} 
\usepackage{lineno} 

\usepackage[normalem]{ulem}
\usepackage{color}
\definecolor{blue}{rgb}{0,0,1}
\definecolor{red}{rgb}{1,0,0}

\journal{Nuclear Physics A} 
\begin{document} 

\begin{frontmatter} 


\title{The NA61/SHINE Experiment at the CERN SPS}

\author{Andr\'as L\'aszl\'o
${}^{a}$, for the NA61/SHINE Collaboration}

\address{
${}^{a}$KFKI Research Inst.\ for Particle and Nucl.\ Physics, Konkoly-Thege
Mikl\'os \'ut 29-33, Budapest, H-1121, Hungary\newline
E-mail address: laszloa@rmki.kfki.hu}

\begin{abstract} 
The physics goals, the detector and its performance as well as 
status and plans of the NA61/SHINE experiment at the CERN SPS accelerator 
are presented.
\end{abstract} 

\end{frontmatter} 



\section{Introduction}\label{introduction}

NA61/SHINE \cite{homepage} is a fixed-target experiment 
at the CERN SPS. The main detector components are inherited from 
the NA49 experiment. These are the two superconducting 
magnets, the four large volume TPCs and two ToF walls.

The physics program of NA61 is the systematic 
measurement of hadron production in proton-proton, 
proton-nucleus, hadron-nucleus, and nucleus-nucleus collisions as a 
function of $\sqrt{s_{{}_{NN}}}$ and beam and target nuclear mass number. 
This comprehensive study has the following main objectives:
(1) search for the critical point by an energy ($E$)- system size ($A$) scan, 
(2) study the properties of the onset of deconfinement by the $E$ - $A$ scan, 
(3) establish, together with the RHIC results, the energy dependence 
    of the nuclear modification factor, 
(4) obtain precision data on hadron spectra in hadron-nucleus collisions for 
    the T2K neutrino experiment, and for the Pierre Auger Observatory and KASCADE 
    cosmic-ray experiments.
In this paper, only the heavy-ion related points (1)-(3) shall be discussed.

\begin{figure}[!ht]
\begin{center}
\includegraphics[width=13cm]{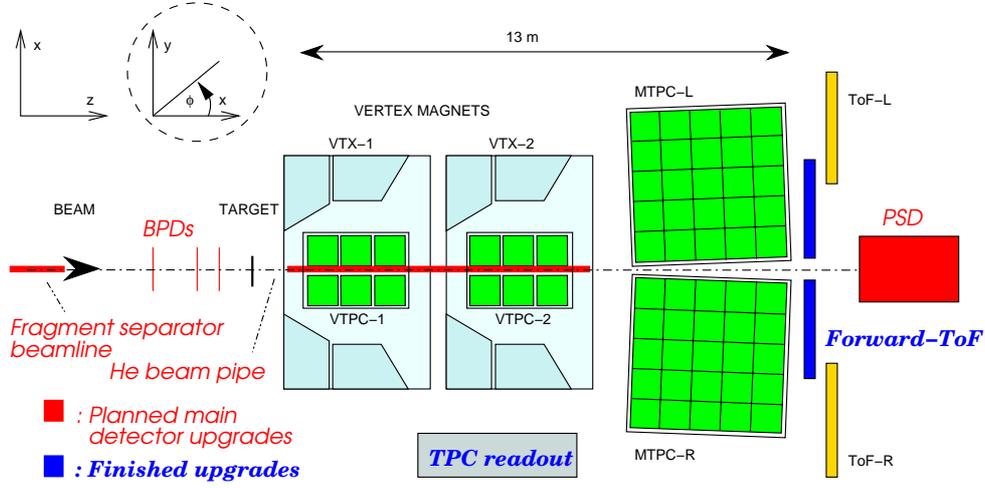}
\end{center}
\caption{The setup of the NA61/SHINE experiment. The main detector 
detector upgrades (accelerated TPC readout, forward ToF, improved 
beam position detectors BPD, new projectile spectator detector PSD,
He filled beam pipe, fragment separator beam line) are shown together 
with the original devices of NA49 (magnets, TPCs, ToF-R/L walls).}
\label{setup}
\end{figure}

To be able to fulfill the demands posed by the physics goals, detector 
upgrades became necessary. The already implemented and the ongoing 
detector improvements are shown in Figure~\ref{setup} and the data taking status 
and plans are listed in Table~\ref{beam}.

\section{Physics goals}

Lattice QCD calculations \cite{katzfodor} indicate that 
the phase diagram of strongly interacting matter features 
a 1-st order phase transition boundary in the temperature - baryochemical 
potential plane, which has a critical endpoint. This critical endpoint 
may be located in the energy range accessible at the CERN SPS.

Temperature ($T$) and baryochemical potential ($\mu_{B}$) of the 
freeze-out points may be scanned via a systematic $E - A$ scan 
\cite{becattini}. Near the critical endpoint an increase of the 
scaled variance ($\omega$) of the multiplicity distribution and of 
the transverse momentum fluctuation measure ($\Phi_{p_{{}_T}}$) are expected 
\cite{rajagopal}. Thus, the critical endpoint may be 
discovered by looking at the energy and system size dependence of 
multiplicity and transverse momentum fluctuations.

To detect an increase of fluctuations related to the critical point, 
the contribution of possible background fluctuations, in particular the 
fluctuation of the number of participant nucleons, have to be minimized. 
As shown in \cite{konchakovski} even for a fixed number of projectile 
participants the number of target participants fluctuates, 
causing a significant background 
in a search for fluctuation signals of the critical point and the 
onset of deconfinement. 
This can be suppressed by selecting very central collisions of identical 
nuclei and by considerable improvement of the spectator energy measurement 
accuracy. 
To achieve this goal, a new projectile spectator energy measurement facility, 
the Projectile Spectator Detector PSD is being built (Figure~\ref{setup}), 
with a resolution of 1 nucleon throughout the energy range of interest.

Another source of background for multiplicity fluctuation measurements 
is the contamination by spiralling low-energy knock-on electrons 
($\delta$-electrons). To minimize this contribution, a Helium beam pipe 
will be introduced around the beam line inside the sensitive TPC volumes 
(Figure~\ref{setup}).

The planned systematic scan in energy and system size will 
allow to study the system size dependence of the anomalies 
in hadron production observed by NA49 \cite{Afanasiev:2002mx} in central Pb+Pb 
collisions at about 30A GeV. These anomalies were predicted 
for the onset of deconfinement \cite{Gazdzicki:1998vd} and their further 
understanding requires new NA61 data.

A very interesting phenomenon, discovered by RHIC experiments at 
$\sqrt{s_{{}_{NN}}}=200\,\mathrm{GeV}$ collision energy, is the reduction 
of high transverse momentum particle yields in nuclear collisions 
relative to elementary collisions (see e.g.\ \cite{adler}), when assuming 
scaling of particle spectra by the number of binary collisions. This 
phenomenon is referred to as `high $p_{{}_T}$ particle suppression', 
and is usually interpreted as the manifestation of parton energy 
loss in the formed strongly interacting matter. Study of the energy 
dependence of the suppression phenomenon is required
for its further understanding. The idea 
is that if the collision energy is low enough such that deconfined
matter is not formed, the high transverse momentum 
particle suppression should disappear.

The recently published low energy $R_{{}_{AA}}$ data on $\pi^{\pm}$ suppression 
at $\sqrt{s_{{}_{NN}}}=17.3\,\mathrm{GeV}$ \cite{highpt} show a monotonic 
increase as a function of $p_{{}_T}$. Unfortunately there are no data points for $p_{{}_T}\geq 2.5\,\mathrm{GeV/c}$. 
Therefore the existence or non-existence of a suppression at higher $p_{{}_T}$ 
is not clear from the present data. The accessible $p_{{}_T}$ range 
was limited by the available p+p statistics of the NA49 experiment. 
The $p_{{}_T}$ range of the reference p+p (and p+Pb) data will be 
extended by high statistics NA61 p+p and p+Pb measurements. 
This program demanded the complete upgrade of the NA61 TPC readout 
system in order to increase the event rate by an order of magnitude.

\section{NA61/SHINE, the upgraded NA49 detector}

A pilot run in 2007 showed that the detector fulfills the physics requirements. 
Its main features are: 
large acceptance ($\approx 50\%$ at $p_{{}_T}\leq2.5\mathrm{GeV/c}$), 
good momentum resolution ($\frac{\sigma(p)}{p^{2}}\approx10^{-4}(\mathrm{GeV/c})^{-1}$), 
good tracking efficiency ($\geq95\%$), 
good particle identification (resolution ToF-L/R: $\sigma(t)\approx60\,\mathrm{ps}$, 
ToF-F: $\sigma(t)\approx120\,\mathrm{ps}$, 
$\frac{\mathrm{d}E}{\mathrm{d}x}$: $\sigma(\frac{\mathrm{d}E}{\mathrm{d}x})\left/\frac{\mathrm{d}E}{\mathrm{d}x}\right.\approx4\%$, 
$V^{0}$ invariant mass: $\sigma(m)\approx5\,\mathrm{MeV}$), 
extended ToF acceptance at low momenta ($\approx1\mathrm{GeV/c}$), 
improved event rate ($70\,\mathrm{Hz}$).

The still progressing upgrades are: 
Projectile Spectator Detector with 1 nucleon precision (prototype tested), 
He beam pipe for reduction of $\delta$-electrons (technical design ready), 
fragment separator for precise selection of secondary ion beams (pilot simulation ready).

\begin{figure}[!ht]
\includegraphics[width=13cm]{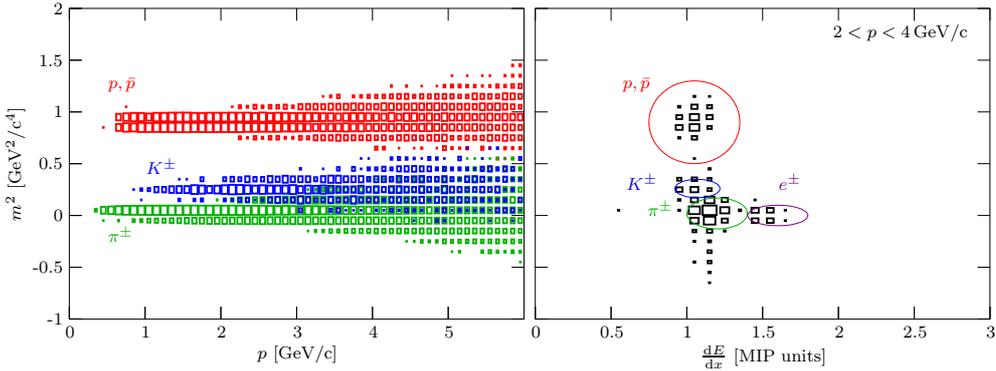}
\caption{Particle identification performance results form the 2007 pilot run 
data. Left: ToF mass-square response as a function of momentum. Right: 
combined ToF + $\frac{\mathrm{d}E}{\mathrm{d}x}$ particle identification in the 
$2<p<4\,\mathrm{GeV/c}$ momentum range. The size of the boxes indicate 
the hit population.}
\label{performance}
\end{figure}

The performance reached in the 2007 pilot run is illustrated in 
Figure~\ref{performance}.

The annual beam requests of NA61 are summarized in Table~\ref{beam}, together 
with the recommendation and approval status assigned by the CERN 
SPS Committee and Research Board.

\begin{table}[!ht]
\begin{center}
\begin{tabular}{llllllll}
 Pri.\ beam             & Beam                       & Target                & E [A GeV]                    & Year              & Days            & Physics                      & Status                         \\
\hline
\hline
              $p$      &             $p$           &             $C$      &             31              &             2007 &             30 &             T2K, CR         &             Run completed     \\
             $p$      &             $p$           &             $C$      &             31              &             2008 &             30 &             T2K, CR         &             Interrupted (LHC) \\
              $p$      &              $p$           &              $C$      &              31              &              2009 &              21 &              T2K, CR         &              Approved          \\
              $p$      &              $\pi^{-}$     &              $C$      &              158, 350        &              2009 &              14 &              CR              &              Approved          \\
              $p$      &              $p$           &              $p$      &              10 - 158        &              2009 &              42 &              CP\&OoD         &              Approved          \\
              $p$      &              $p$           &              $p$      &              158             &              2010 &              77 &              high $p_{{}_T}$ &              Recommended       \\
              $Pb$     &              $S$           &              $S$      &              10 - 158        &              2011 &              42 &              CP\&OoD         &              Recommended{          ${}^{*}$} \\
              $p$      &              $p$           &              $Pb$     &              158             &              2011 &              42 &              high $p_{{}_T}$ &              Recommended       \\
              $p$      &              $p$           &              $Pb$     &              10 - 158        &              2012 &              42 &              CP\&OoD         &              Recommended       \\
           $Pb$     &           $C$           &           $C$      &           10 - 158        &           2012 &           42 &           CP\&OoD         &           To be discussed${}^{*}$ \\
           $Pb$     &           $In$          &           $In$     &           10 - 158        &           2013 &           42 &           CP\&OoD         &           To be discussed${}^{*}$ \\
\hline
\end{tabular}
\end{center}
\caption{The beam request of the NA61/SHINE experiment. Abbreviations: CP -- search for Critical Point; 
OoD -- study the Onset of Deconfinement; T2K -- supplementary spectra for the T2K experiment; 
CR -- measurements for cosmic-ray physics; high $p_{{}_T}$ -- p+p and p+Pb reference spectra 
for nuclear modification factors. ${}^{*}$: Needs implementation of fragment separator beam line 
as the LHC beam schedule only allows Pb as primary heavy-ion beam in the SPS.}
\label{beam}
\end{table}

For the approval procedure of the NA61 experiment, the following documents are 
the most relevant: 
Expression of Interest \cite{eoi}, Letter of Intent \cite{loi}, Status Report 
\cite{sr}, Proposal \cite{prop}, Addendum-1 \cite{add1} and Addendeum-2 \cite{add2}.

\section{Summary}

The experiment NA61/SHINE has great discovery potential for 
the critical point of strongly interacting matter, if it exists.

Important measurements of the nuclear modification factor 
at the top SPS energy can be performed and the system size dependence 
of the effects related to the onset of deconfinement can be studied.

NA61 will provide necessary 
supporting measurements of hadron production 
for neutrino and cosmic-ray experiments.

The pilot run in 2007 has been successfully completed, and shows that the 
detector performance is sufficiently good. The $E - A$ scan starts in 2009 
with p+p collisions. Heavy-ion measurements will start in 2011. There 
are also further projects on nucleus-nucleus collisions in the SPS energy range 
currently developed at BNL, FAIR and NICA, addressing the discussed 
physics questions.





\begin{thebibliography}{00} 
\bibitem{homepage} \emph{The NA61/SHINE homepage} [\texttt{http://na61.web.cern.ch}].
\bibitem{katzfodor} Z.~Fodor, S.~D.~Katz, \emph{JHEP}~{\bf 0203} (2002) 014.
\bibitem{becattini} F.~Becattini, J.~Manninen, M.~Gazdzicki, \emph{Phys.~Rev.}~{\bf C73} (2006) 044905.
\bibitem{rajagopal} M.~Stephanov, K.~Rajagopal, E.~Shuryak, \emph{Phys.~Rev.}~{\bf D60} (1999) 114028.
\bibitem{konchakovski} V.~P.~Konchakovski et al, \emph{Phys.~Rev.}~{\bf C73} (2006) 034902.
\bibitem{Afanasiev:2002mx} 
 S.~V.~Afanasiev et al (the NA49 Collaboration), 
 Phys.~Rev.~{\bf C66} (2002) 054902.
\bibitem{Gazdzicki:1998vd} M.~Gazdzicki, M.~I.~Gorenstein, 
 \emph{Acta~Phys.~Polon.}~{\bf B30} (1999) 2705.
\bibitem{adler} S.~S.~Adler et al (the PHENIX Collaboration), \emph{Phys.~Rev.~Lett.}~{\bf 91} (2003) 072303.
\bibitem{highpt} C.~Alt et al (the NA49 Collaboration), \emph{Phys.~Rev.}~{\bf C77} (2008) 034906.
\bibitem{eoi} J.~Bartke et al (the NA61 Collaboration), \emph{A new experimental programme with nuclei and proton beams at the CERN SPS}, NA49-future expression of interest (2003), CERN-SPSC-2003-031, SPSC-EOI-001.
\bibitem{loi} N.~Antoniou et al (the NA61 Collaboration), \emph{Study of hadron production in collisions of protons and nuclei at the CERN SPS}, NA49-future letter of intent (2006), CERN-SPSC-2006-001, SPSC-I-235.
\bibitem{sr} N.~Antoniou et al (the NA61 Collaboration), \emph{Report from tests of the NA49 experimental facility and the NA49-future detector prototypes}, NA49-future status report (2006), CERN-SPSC-2006-023, SPSC-SR-010.
\bibitem{prop} N.~Antoniou et al (the NA61 Collaboration), \emph{Study of hadron production in hadron-nucleus and nucleus-nucleus collisions at the CERN SPS}, NA49-future proposal (2006), CERN-SPSC-2006-034, SPSC-P-330.
\bibitem{add1} N.~Antoniou et al (the NA61 Collaboration), \emph{Additional information requested in the proposal review process}, Addendum-1 to the NA49-future proposal (2007), CERN-SPSC-2007-004, SPSC-P-330.
\bibitem{add2} N.~Antoniou et al (the NA61 Collaboration), \emph{Further information requested in the proposal review process}, Addendum-2 to the NA49-future proposal (2007), CERN-SPSC-2007-019, SPSC-P-330.
\end{thebibliography}
\end{document}